\documentstyle[sprocl]{article}

\def\bull{\parskip=0pt\par\noindent\hangindent=3pc\hangafter=1 $\bullet$~}

\def\kms{\ifmmode{\,\hbox{km}\,s^{-1}}\else {\rm\,km\,s$^{-1}$}\fi}

\def\hmpc{\ifmmode{h^{-1}\,\hbox{Mpc}}\else{$h^{-1}$\thinspace Mpc}\fi}

\def\sigp{\ifmmode{\sigma_p}\else {$\sigma_p$}\fi}
\def\sig1{\ifmmode{\sigma_1}\else {$\sigma_1$}\fi}
\def\r200{\ifmmode{r_{200}}\else {$r_{200}$}\fi}
\def\spose#1{\hbox to 0pt{#1\hss}}
\def\lta{\mathrel{\spose{\lower 3pt\hbox{$\mathchar"218$}}
     \raise 2.0pt\hbox{$\mathchar"13C$}}}
\def\gta{\mathrel{\spose{\lower 3pt\hbox{$\mathchar"218$}}
     \raise 2.0pt\hbox{$\mathchar"13E$}}}

\def\apj{ApJ}
\def\apjl{ApJ(Lett)}

\def\apjs{ApJS}

\def\aj{AJ}

\input{epsf}
\input{rotate}

\begin{document}

\title{$\Omega_M$ AND THE CNOC SURVEYS }

\author{ R.G. CARLBERG, H.K.C. YEE, H. LIN, M. SAWICKI, C.W. SHEPHERD,}
\address{Department of Astronomy, University of Toronto, 
	Toronto, ON, M5S 3H8, Canada}

\author{E. ELLINGSON}
\address{Center for Astrophysics \& Space Astronomy,
        University of Colorado, CO 80309, USA}

\author{S.L. MORRIS, D. SCHADE, J.E. HESSER, J. B. HUTCHINGS, J. B. OKE}
\address{National Research Council of Canada,
        Herzberg Institute of Astrophysics,
        Dominion Astrophysical Observatory, 
        Victoria, BC, V8X~4M6, Canada}

\author{D. PATTON, G. WIRTH, M. BALOGH, F.D.A. HARTWICK, C. J. PRITCHET}
\address{Department of Physics and Astronomy, University of Victoria,
        Victoria, BC, V8W~3P6, Canada}

\author{R. ABRAHAM}
\address{Institute of Astronomy, 
        Madingley Road, Cambridge CB3~0HA, UK}

\author{T. SMECKER-HANE}
\address{Department of Physics \& Astronomy,
        University of California, Irvine,
        CA 92717, USA}


\maketitle\abstracts{ The CNOC1 cluster survey measures $\Omega_M$ via
Oort's method, $\Omega_M\equiv M/L \times j/\rho_c$, where $M/L$ is the
field mass-to-light ratio, $j$ is the field luminosity density and
$\rho_c$ is the closure density.  A wide range of potential systematic
effects are explicitly controlled by independently deriving the mean
cluster mass profile (finding good agreement with theoretical
predictions), the cluster light profile, the redshift evolution of
both cluster and field galaxies, the differential evolution between
the two, and the field and cluster efficiencies for the conversion of
baryons into galaxies.  We conclude that $\Omega_M=0.19\pm0.06$ where
the errors are objectively evaluated via resampling methods.  The
redshift evolution of the numbers of clusters per unit co-moving
volume over the $0\le z \le 0.6$ range is found to be very slow, as is
required for consistency with a low density universe.  The evolution
of galaxy clustering in the field is compatible with a low density
universe, and strongly disfavors models of galaxy evolution that
associate low density halos with individual galaxies.}
  
\section{The Mean Mass Density of the Universe}

In the Friedmann-Robertson-Walker solution for the structure of the
universe the geometry and future of the expansion uniquely depend on
the mean mass density, $\rho_0$, and a non-zero cosmological constant.
It is a statement of arithmetic\,\cite{oort} that $\Omega_M
\equiv\rho_0/\rho_c= M/L
\times j/\rho_c$, where $M/L$ is the average mass-to-light ratio of
the universe and $\rho_c/j$ is the closure mass-to-light ratio, with
$j$ being the luminosity density of the universe.  Estimates of the
value of $\Omega_M$ have a long history with a substantial range of
cited results\,\cite{bld}. Both the ``Dicke coincidence'' and
inflationary cosmology would suggest that $\Omega_M=1$.  The main
thrust of our survey is to clearly discriminate between $\Omega_M=1$
and the classical, possibly biased, indicators that $\Omega_M\simeq
0.2$.

Rich galaxy clusters
are the largest collapsed regions in the universe and are ideal to
make an estimate of the cluster $M/L$ which can be corrected to the
value which should apply to the field as a whole.  To use clusters to
estimate self-consistently the global $\Omega_M$ we must, as a
minimum, perform four operations.
\bull{Measure the total gravitational mass within some radius.}
\bull{Sum the luminosities of the visible galaxies within the same
	radius.}
\bull{Measure the luminosity density in the field of an identically defined
	galaxy sample.}
\bull{Account (and ideally physically understand) the 
	differential luminosity and density evolution between
	the clusters and the field.}

The Canadian Network for Observational Cosmology (CNOC) designed
observations to make a conclusive measurement of $\Omega_M$ using
clusters.  The clusters are selected from the X-ray surveys, primarily
the Einstein Medium Sensitivity Survey\,\cite{emss1,emss2,gl}, which
has a well defined flux-volume relation. The spectroscopic sample,
roughly one in two on the average, is drawn from a photometric sample
which goes nearly 2 magnitudes deeper, thereby allowing an accurate
measurement of the selection function.  The sample contains 16
clusters spread from redshift 0.18 to 0.55, meaning that evolutionary
effects are readily visible, and any mistakes in differential
corrections should be more readily detectable.  For each cluster,
galaxies are sampled all the way from cluster cores to the distant
field. This allows testing the accuracy of the virial mass estimator
and the understanding of the differential evolution process.  We
introduce some improvements to the classical estimates of the velocity
dispersion and virial radius estimators, which have somewhat better
statistical properties.  A critical element is to assess the errors in
these measurements.  The random errors are relatively straightforward
and are evaluated using either the statistical jacknife or bootstrap
methods\,\cite{et}.  These resampling methods are completely objective
and follow the entire complex chain of analysis beginning from the
input catalogues to the result.

As a result of these measurements and tests we find
that $\Omega_M=0.19\pm0.06$, which is the formal $1\sigma$ error.
In deriving this result we apply a variety of corrections
and tests of the assumptions.
\bull{The clusters have statistically identical $M/L$ values,
	once corrected for evolution\,\cite{global}.}
\bull{Cluster and field galaxies are evolving at a comparable
	rate with redshift, approximately one magnitude per
	unit redshift.}
\bull{Cluster galaxies exhibit no excess in their star formation
	with respect to the field\,\cite{a2390,balogh}. On the average they
	are faded between 0.1 and 0.3 magnitudes with respect
	to luminous field galaxies\,\cite{profile,lin}.}
\bull{The virial mass overestimates the true mass of a cluster
	by about 15\%, which can be attributed to the neglect
	of the surface term in the virial equation\,\cite{profile}.}
\bull{There is no significant change of $M/L$ with radius
	within the cluster\,\cite{profile}.}
\bull{The mass field of the clusters is remarkably well described
	by the NFW\,\cite{nfw} profile, both in shape and scale 
	radius\,\cite{ave}.}
\bull{The evolution of the number of clusters per unit volume is
	very slow, in accord with the PS\,\cite{ps} predictions for
	a low density universe\,\cite{s8}.}
\bull{The clusters have statistically identical efficiencies of
	converting gas into stars, which is identical to the
	value in the field\,\cite{omb}.}

These results rule out $\Omega_M=1$ in any component with a velocity
dispersion less than about 1000 \kms.

\section{The Future of Clusters as Cosmological Indicators}

The dark matter distribution within clusters is now quite well
understood and the differential of field and cluster galaxies is
becoming well observed and to some degree understood. With much larger
samples it will be possible to more tightly constrain many
cosmological quantities.  Of particular interest is that
$\Omega_\Lambda$ can be measured via the redshift dependence of the
$M/L\times j/\rho_c$ indicator, being nearly a 50\% change from
redshift zero to unity. The main complication is to make sure that
differential field-to-cluster evolution is accurately measured and
physically well understood at a somewhat better level of
precision. Statistically this is not a problem, since the survey will
contain sufficient galaxies in both cluster and field to make the
measurement. Straightforward simulations show\,\cite{lambda} that it
is statistically possible to measure $\Omega_\Lambda$ with a sample of
30 or so clusters distributed over the $0\le z \le 1$ range.  With a
sample of 200 to 300 clusters over this range it will be possible to
measure the $\Omega$ parameters to an accuracy of better than 10\%, as
well as developing an impressive sample of both field and cluster
galaxies.

\section{Slow Structure evolution for $\Omega_M=0.2$}

A low density universe ``freezes out'' structure at
redshift\,\cite{lss} $z\simeq \Omega_M^{-1}$ with relatively slow
growth in clustering after that. This allows a test of both the value
of $\Omega_M$ and the understanding of the relation between the
clustering of dark matter and the galaxies that we observe.  The
measurements of clustering can be conveniently approximated at a level
appropriate for existing data by a double power law\,\cite{gp,ks} in
pairwise separation (measured in physical or proper lengths) and in
$1+z$.
\begin{equation}
\xi(r|z)=\left({r\over r_0}\right)^{-\gamma} (1+z)^{-(3+\epsilon)}.
\label{eq:def}
\end{equation}
The $(1+z)^{-3}$ dilution of the correlation function is simply
a result of the change of the background density of the universe
with redshift. At low redshift\,\cite{dp,apm} $\gamma\simeq 1.7-1.8$.
Very crudely, there are three ``interesting'' value
of $\epsilon$. 
\bull{$\epsilon\simeq 0$ for a low density, e.g. $\Omega_M\simeq0.2$, 
	universe,}
\bull{$\epsilon\simeq -1$ for a high density, $\Omega_M=1$, universe,}
\bull{$\epsilon\simeq 1$ for low overdensity, $\langle \rho \rangle 
	\simeq 200 \rho_c$, dark halos.}  

\noindent
The last possibility is interesting since a number of
``semi-analytic'' models of galaxy formation identify galaxies with
these objects, partly because they can be counted using the
Press-Schechter formula. The correlation properties of halos depend on
the details\,\cite{cbgxi,bv}, but to a rough approximation the low
overdensity halos cluster with a correlation length that is fixed in
co-moving co-ordinates.

\medskip
\vbox{
\hbox{
\epsfysize 5.5truecm
\epsfbox[40 200 612 700]{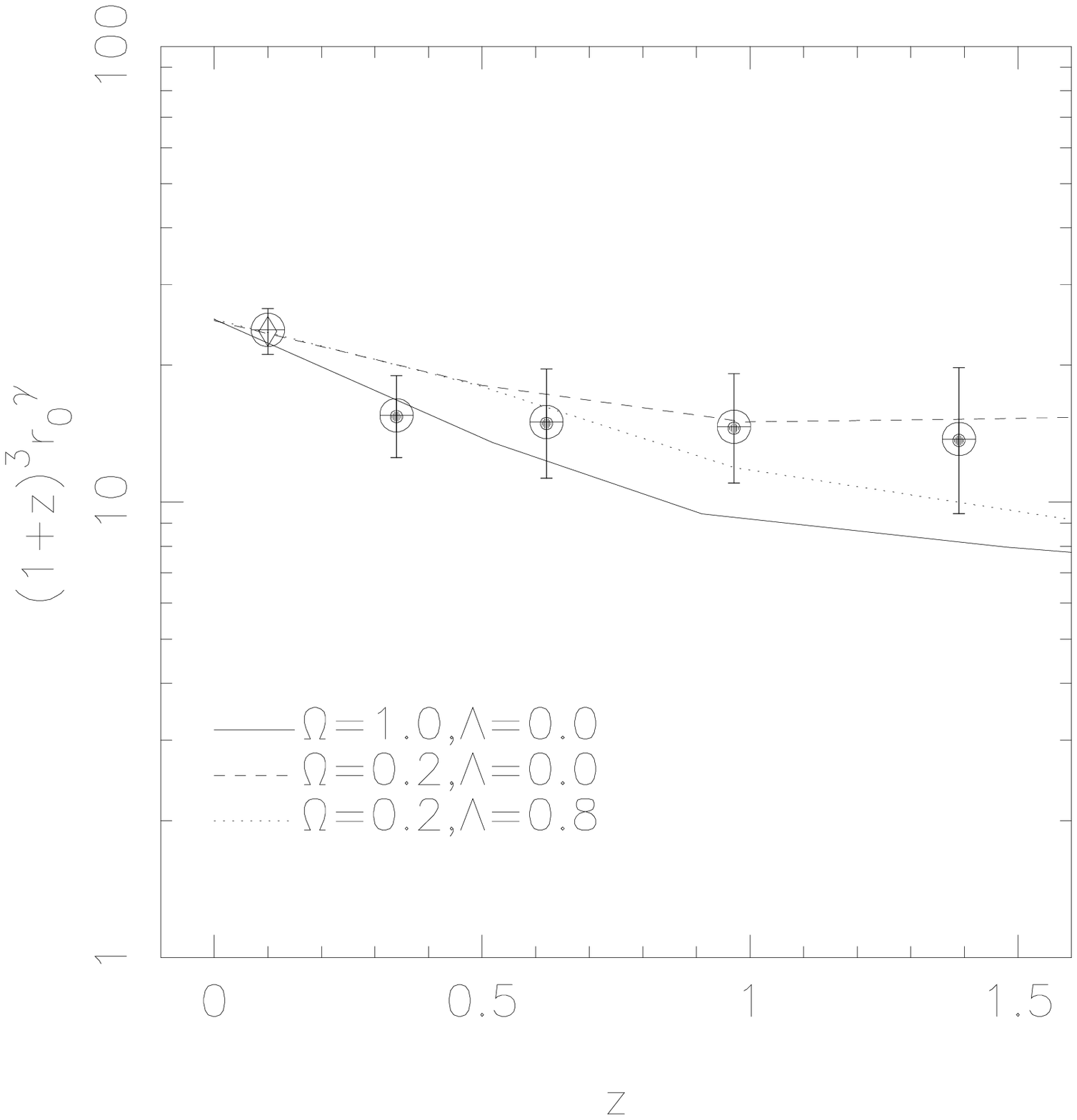}
~\epsfysize 5.5truecm
\epsfbox[40 200 612 700]{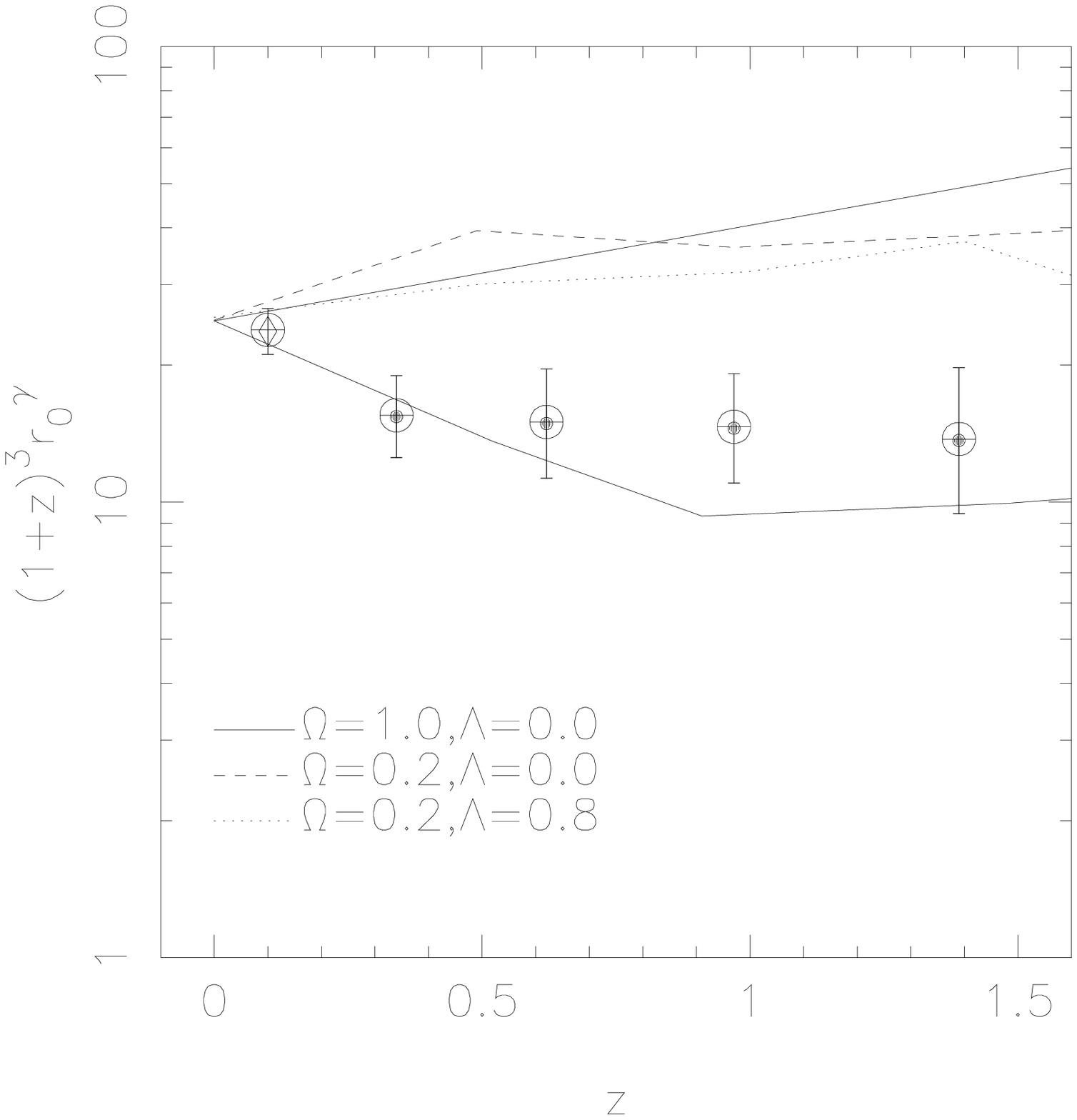}
}
\noindent
\footnotesize Figure 1: The evolution of measured correlations of red
selected galaxies, R band for LCRS at $z=0.1$ and the Hawaii K sample
at higher redshifts. The lines show the predicted evolution for the
full mass field (left) and the $2000\rho_0$ (not $200\rho_c$) halos on
the right and the $\epsilon=+1.2$ ``fixed in co-moving co-ordinates''
clustering as the solid straight line.  }
\medskip
The predictions\,\cite{ccc} for three model universes,
$[\Omega_M,\Omega_\Lambda]$ of $[1,0]$, $[0.2,0]$ and $[0.2,0.8]$, are
compared to R band selected data from the Las Campanas Redshift
Survey\,\cite{tucker} and the Hawaii K-band survey\,\cite{cshc,kcor}
in Figure~1. Predictions are made for both ``galaxies-trace-mass'' and
for ``dark halos-trace-mass'' galaxy identification schemes.  Red
selected galaxies are used because they are found to be more reliable
tracers of the mass distribution in the cluster survey, as expected on
the basis of a reduced sensitivity in the redder pass bands to the
history of star formation. However in detail the question of how
galaxies trace the mass distribution remains open to investigation and
is a major motivating factor in the CNOC2 survey of galaxies in the
field.  The major conclusion to be drawn from Figure~1 is that low
overdensity halos, or any other object which clusters in such a way
that its correlations remain roughly constant in co-moving
co-ordinates, are in conflict with the available measurements of
clustering evolution.

\section{The CNOC2 Field Survey}

\smallskip
\vbox{
\hbox{
\epsfysize 5.5truecm
\epsfbox[20 200 612 750]{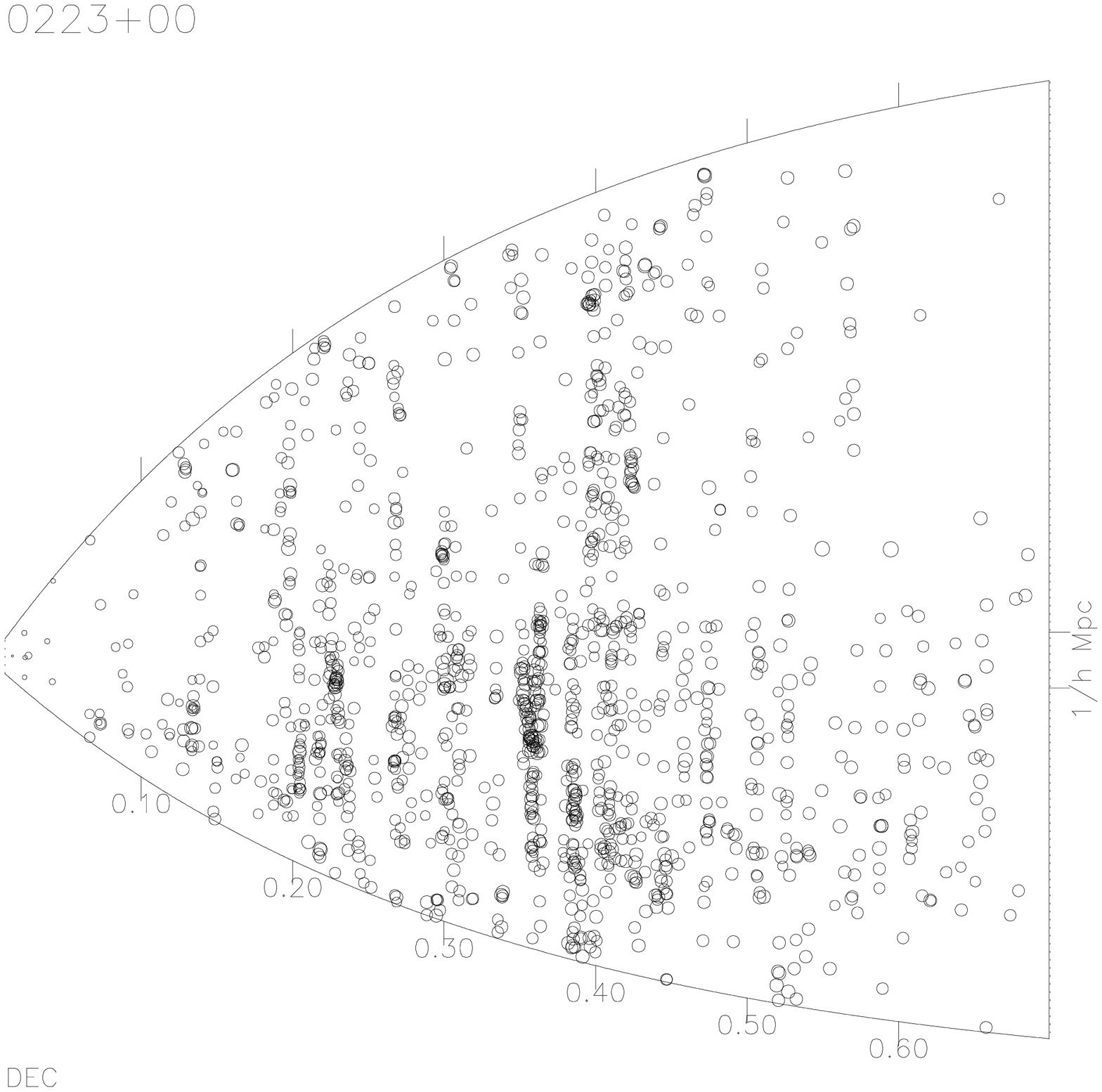}
~\epsfysize 5.5truecm
\epsfbox[20 200 612 750]{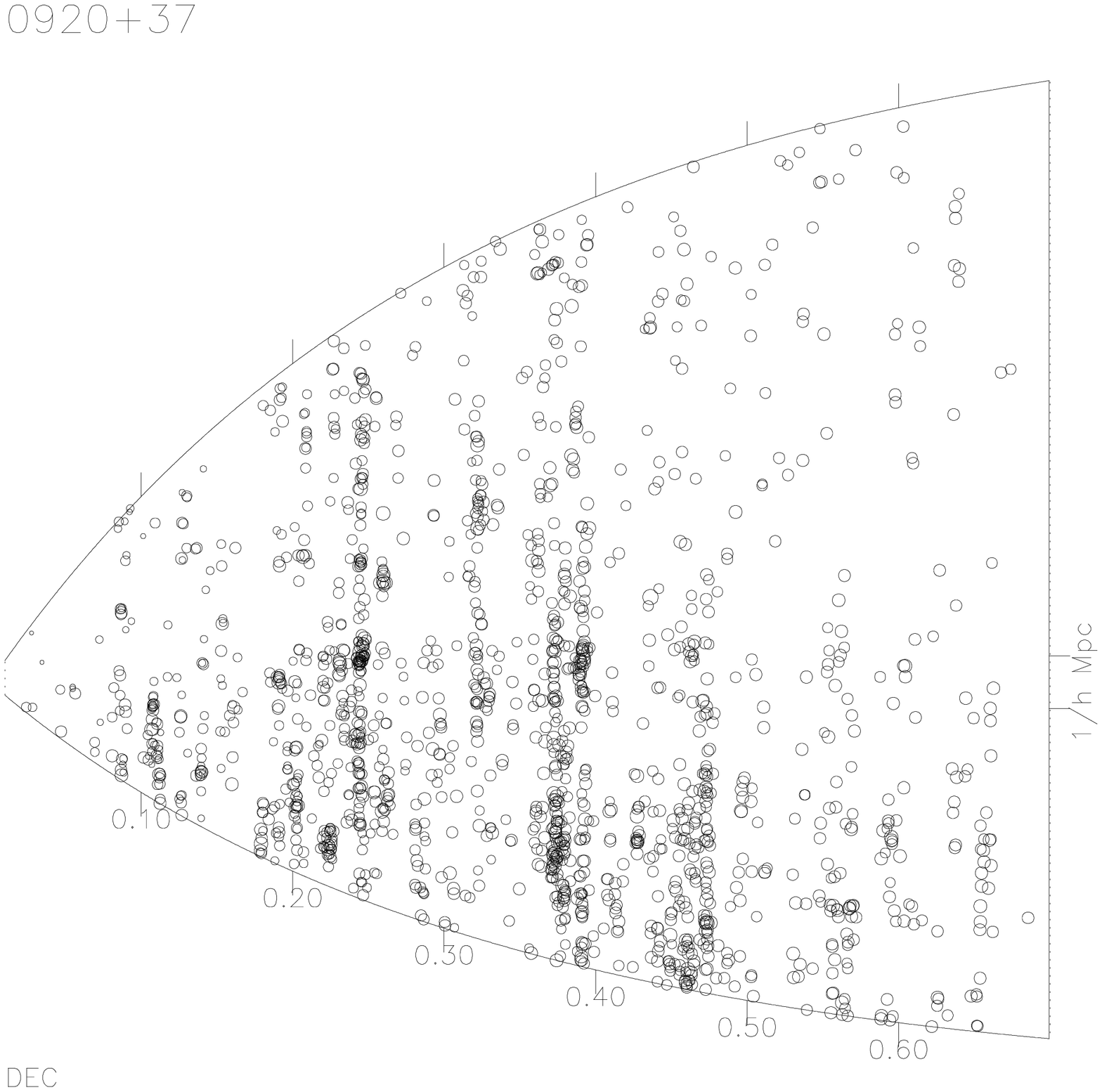}
}
\noindent
\footnotesize
Figure 2: The physical distance in the Dec direction from the field
center plotted against redshift in two of the CNOC2 fields. Note the
strong large scale clustering.  The survey has a greater RA extent at
smaller Dec (the patches are roughly ``L'' shaped) hence the
greater density of points at the bottom of the pies.}
\medskip

The CNOC2 field survey is designed to study the dynamics and clustering
of faint galaxies and its relation to galaxy properties. The survey
is statistically complete to Cousins $R=21.5$ mag and $B=22.5$
mag. The survey covers approximately 1.5 square degrees of sky, in 4
patches having approximately an ``L'' shape with the largest
separations being 1.5$^\circ$, and a central block of 0.5$^\circ$ on a
side. The survey is now about 75\% complete and contains more than
5000 galaxies with redshifts accurate to about 70 \kms\ in the rest
frame.  In addition there are more than 20,000 galaxies, about 1 mag
deeper, with accurate UBgRI photometry, which is essential for
measuring the selection function and understanding galaxy evolution.

\section*{Acknowledgments}
This research was supported by NSERC and NRC of Canada. We thank
the Canadian TAC of CFHT for a generous allocation of telescope time.

\end{document}